\documentclass[prl,preprint,endfloats]{revtex4}

\usepackage{graphicx}
\usepackage{bm}
\usepackage{pifont}
\usepackage{amsmath}

\begin{document}

\title{Spontaneous topological Hall effect induced by non-coplanar antiferromagnetic order in intercalated van der Waals materials}

\author{H. Takagi$^{1}$, R. Takagi$^{1,2,3,4}$, S. Minami$^{5}$, T. Nomoto$^{6}$, K. Ohishi$^{7}$, M.-T. Suzuki$^{8,9}$, Y. Yanagi$^{8,10}$, M. Hirayama$^{1,4}$, N. D. Khanh$^{4}$, K. Karube$^{4}$, H. Saito$^{11}$, D. Hashizume$^{4}$, R. Kiyanagi$^{12}$, Y. Tokura$^{1, 4, 13}$, R. Arita$^{4,6}$, T. Nakajima$^{4,11}$, S. Seki$^{1,2,3,4}$}

\affiliation{$^1$ Department of Applied Physics, University of Tokyo, Tokyo 113-8656, Japan, \\ $^2$ Institute of Engineering Innovation, University of Tokyo, Tokyo 113-8656, Japan, \\ $^3$ PRESTO, Japan Science and Technology Agency (JST), Kawaguchi 332-0012, Japan, \\ $^4$ RIKEN Center for Emergent Matter Science (CEMS), Wako 351-0198, Japan, \\ $^5$ Department of Physics, University of Tokyo, Tokyo 113-8656, Japan, \\ $^6$ Research Center for Advanced Science and Technology, University of Tokyo, Tokyo 153-8904, Japan, \\ $^7$ Neutron Science and Technology Center, Comprehensive Research Organization for Science and Society (CROSS), Tokai 319-1106, Japan, \\ $^8$ Center for Computational Materials Science, Institute for Materials Research, Tohoku University, Sendai 980-8577, Japan, \\ $^9$ Center for Spintronics Research Network, Graduate School of Engineering Science, Osaka University, Toyonaka, Osaka 560-8531, Japan, \\ $^{10}$ Liberal Arts and Sciences, Toyama Prefectural University, Imizu, Toyama 939-0398, Japan, \\ $^{11}$ The Institute for Solid State Physics, University of Tokyo, Kashiwa, Chiba, Japan, \\ $^{12}$ J-PARC Center, Japan Atomic Energy Agency, Tokai, Naka-gun 319-1195, Japan, \\ $^{13}$ Tokyo College, University of Tokyo, Tokyo 113-8656, Japan.}

\begin{abstract}

{\bf In ferromagnets, electric current generally induces transverse Hall voltage in proportion to magnetization (anomalous Hall effect), and it is frequently used for electrical readout of the $\uparrow$ and $\downarrow$ spin states. While these properties are usually not expected in antiferromagnets, recent theoretical studies predicted that non-coplanar antiferromagnetic order with finite scalar spin chirality (i.e. solid angle spanned by neighboring spins) can often induce large spontaneous Hall effect even without net magnetization or external magnetic field. This phenomenon, i.e. spontaneous topological Hall effect, can potentially be used for the efficient electrical readout of the antiferromagnetic states, but its experimental verification has long been elusive due to the lack of appropriate materials hosting such exotic magnetism. Here, we report the discovery of all-in-all-out type non-coplanar antiferromagnetic order in triangular lattice compounds Co$M_3$S$_6$ ($M=$ Nb, Ta), by performing the detailed magnetic structure analysis based on polarized neutron scattering experiments as well as systematic first-principles calculations. These compounds are reported to host unconventionally large spontaneous Hall effect despite their vanishingly small net magnetization, and our analysis revealed that it can be well explained in terms of topological Hall effect, which originates from the fictitious magnetic field associated with scalar spin chirality in non-coplanar antiferromagnetic orders. The present results indicate that the scalar spin chirality mechanism can offer a promising route to realize giant spontaneous Hall response even in compensated antiferromagnets, and highlight intercalated van der Waals magnets as an unique quasi-two-dimensional material platform to enable various nontrivial manner of electrical reading and possible writing of non-coplanar antiferromagnetic domains.}

\end{abstract}
\maketitle

Today, magnetic information is mostly stored by the $\uparrow$ and $\downarrow$ spin states in ferromagnets. They are transformed into each other by time-reversal operation, and can be distinguished due to time-reversal symmetry (TRS) breaking. In these two states, electric current generally induces the opposite sign of transverse Hall voltage in proportion to magnetization $M$ (anomalous Hall effect (Fig. 1a)), and this phenomenon is frequently used for the electrical readout of the $\uparrow$ and $\downarrow$ spin states\cite{RMP_AHE}. In contrast, such functions are usually not expected in antiferromagnets, because of their time-reversal symmetry and lack of macroscopic magnetization.

Recently, however, a few antiferromagnets with nontrivial spin texture were reported to host large spontaneous Hall effect even without net magnetization or external magnetic field $H$. One typical example is Kagome lattice antiferromagnet Mn$_3$Sn\cite{Mn3Sn_MacDonald_Theory, Mn3Sn_Nature, Mn3Ge, Mn3Sn_ARPES, Mn3Sn_Kerr, Mn3Sn_Econtrol}. In this compound, the coplanar $120^\circ$ magnetic structure breaks time-reversal symmetry, and the spin-orbit coupling and associated band splitting are considered as the main source of large spontaneous Hall signal\cite{Mn3Sn_MacDonald_Theory, Mn3Sn_ARPES}. Such TRS-broken antiferromagnets are expected to show similar functional response as ferromagnets\cite{Mn3Sn_Nature, Mn3Sn_Kerr, Mn3Sn_Econtrol}, and now intensively studied as a candidate of novel information medium with various unique advantages such as negligibly small stray field, faster spin dynamics, and stability against magnetic perturbation\cite{RMP_Ono}. So far, the number of this class of materials is very limited, and further search of exotic TRS-broken antiferromagnetic spin texture, as well as the novel microscopic mechanism to realize large spontaneous Hall response, is highly demanded.

Our target materials, triangular lattice compounds Co$M_3$S$_6$ ($M$ = Nb, Ta)\cite{ParkinMagnetization, ParkinTransport, ParkinNeutron}, are another candidate of such TRS-broken antiferromagnet. These compounds were recently reported to host unconventionally large spontaneous Hall effect despite their negligibly small net magnetization\cite{CoNb3S6_AHE, CoTa3S6_AHE}, while its microscopic origin and associated magnetic structure are still in controversy. The previous unpolarized neutron scattering experiment suggested collinear or coplanar (non-collinear) antiferromagnetic orders\cite{ParkinNeutron}, and the most of existing works rely on this report\cite{CoNb3S6_AHE, CoTa3S6_AHE, CrystalHall}. On the other hand, the latest theoretical studies proposed the possible appearance of non-coplanar antiferromagnetic order\cite{CoNb3S6_noncoplanar_Calc}, but this scenario has not been verified experimentally.

Here, the latter scenario is particularly interesting, because non-coplanar spin texture is theoretically predicted to cause topological Hall effect (Fig. 1b), which originates from geometrical character of real-space spin texture without the need of spin-orbit coupling\cite{RMP_AHE, Nagaosa_fcc, Batista, Motome_SSC_JPSJ, Blugel_TopologicalKerr_fcc}. In general, conduction electrons passing through non-coplanar spin texture pick up additional quantum-mechanical Berry phase, and feel a local fictitious magnetic field in proportion to the scalar spin chirality $\chi_{ijk} \equiv {\bf S}_i \cdot ({\bf S}_j \times {\bf S}_k)$, i.e., the solid angle spanned by neighboring spins ${\bf S}_i$, ${\bf S}_j$, and ${\bf S}_k$ \cite{Taguchi, Nagaosa_fcc, RMP_AHE}. It leads to the emergence of topological Hall effect (Fig. 1b), whose amplitude scales with the fictitious magnetic field rather than $M$. Previously, such a topological Hall effect arising from scalar spin chirality mechanism has been studied in several non-coplanar ferromagnets with sizable macroscopic magnetization\cite{Taguchi, MnSi_THE, Gd2PdSi3_THE}. Nevertheless, this phenomenon has rarely been identified in compensated antiferromagnets experimentally, partly because their fictitious magnetic field frequently cancels out by the symmetry\cite{Nagaosa_fcc} and also the non-coplanar antiferromagnetic order itself is uncommon. Since this scalar spin chirality mechanism can potentially offer a novel route to realize large spontaneous Hall response in antiferromagnets, its verification in the real material is an important challenge.

In this study, we report the experimental discovery of all-in-all-out type non-coplanar antiferromagnetic order in triangular lattice compounds Co$M_3$S$_6$ ($M=$ Nb, Ta), by performing the detailed magnetic structure analysis based on polarized neutron scattering measurements. It clarifies the microscopic origin of observed unconventionally large spontaneous Hall signal with vanishingly small net magnetization, which can be well explained in terms of topological Hall effect. The present results demonstrate that the scalar spin chirality mechanism can be another promising route to realize giant spontaneous Hall response in compensated antiferromagnets, and highlight intercalated van der Waals magnets as an unique quasi-two-dimensional material platform to enable various nontrivial manner of electrical reading and possible writing of non-coplanar antiferromagnetic domains.

Figure 1c indicates the crystal structure of our target compound Co$M_{3}$S$_{6}$ ($M=$ Nb, Ta). They can be considered as a member of intercalated transition metal dichalcogenides, where magnetic Co ions are inserted into van der Waals gaps of the parent two-dimensional compound 2H-$M$S$_2$\cite{ParkinMagnetization, ParkinTransport}. The intercalated Co ions form triangular lattice layers, which can be viewed as $\sqrt{3}\times\sqrt{3}$ superstructures with respect to the $M$S$_2$ triangular lattice layers. It leads to the breaking of space-inversion and mirror symmetries, and the overall crystal structure is characterized by the noncentrosymmetric chiral hexagonal space group $P6_3 22$. In the following, we mainly focus on CoTa$_3$S$_6$ for simplicity, while similar experimental data and conclusions are also obtained for CoNb$_3$S$_6$ as detailed in Supplementary Note IV.

Figure 1d summarizes the $H$-$T$ (temperature) magnetic phase diagram of CoTa$_3$S$_6$ for $H \parallel c$ (See Supplementary Note I and Supplementary Fig. 1 for the detail). The antiferromagnetic order appears below N\'{e}el temperature $T_{{\rm N1}}$ = 38 {\rm K}, and additional magnetic transition is identified at $T_{{\rm N2}}$ = 24 {\rm K}. Figures 1e-g indicate the magnetic field dependence of magnetization $M$, longitudinal resistivity $\rho_{xx}$ and Hall resistivity $\rho_{yx}$ measured for $H\parallel c$ at 10 {\rm K}. The magnetization profile shows a clear hysteresis loop with tiny spontaneous magnetization $\Delta M \sim 0.009$ $\mu_{\rm B}/$Co at $H=0$ (Phase I), and the sign of $\Delta M$ is reversed at 3 {\rm T}. In addition, a clear step-like anomaly is also found at 4.5 {\rm T}, suggesting the emergence of the additional metamagnetic transition into Phase II. In the corresponding $\rho_{\mathrm{yx}}$ profile (Fig. 1g), large spontaneous Hall signal can be identified at zero field. It is characterized by a hysteresis loop with the sign reversal at 3 T, and additional step-like anomaly is also identified at 4.5 {\rm T}. These results indicate the strong correlation between the magnetism and electron transport properties, in accord with the previous report\cite{CoTa3S6_AHE}. Throughout this process, $\rho_{\mathrm{xx}}$ remains almost constant (Fig. 1f). In conventional ferromagnets, $\rho_{\mathrm{yx}}$ is generally described by $\rho_{\mathrm{yx}}=R_{{\rm 0}}H+R_{\mathrm{s}}M$, where the first and second terms represent the normal and anomalous Hall term proportional to $H$ and $M$, respectively ($R_0$ and $R_{\mathrm{s}}$ are the coefficients for each term)\cite{RMP_AHE}. In the present case of CoTa$_3$S$_6$, however, the experimental $\rho_{\mathrm{yx}}$ profile (Fig. 1g) is clearly not proportional to $M$ (Fig. 1e). In addition, the observed giant spontaneous Hall angle $\rho_{\mathrm{yx}} (H=0) / \rho_{\mathrm{xx}}(H=0) \sim 2$ {\rm \%} is comparable with, or even larger than, typical ferromagnets, which cannot be explained by the tiny spontaneous magnetization $\Delta M$ (See Supplementary Note I for the detail). The above results strongly suggest that the spontaneous Hall effect in CoTa$_{3}$S$_{6}$ emerges from a nontrivial origin, rather than the conventional $M$-linear anomalous Hall contribution.

To identify its microscopic origin, we performed the magnetic structure analysis based on neutron scattering measurements for a CoTa$_3$S$_6$ single crystal. Figure 2a indicates the schematic illustration of magnetic and nuclear reflections in the reciprocal space observed in the Phase I at $H=0$. The corresponding ($\delta$, 0, 0) and ($\delta$, $\delta$, 0) line scan profiles measured at 0 {\rm T} are shown in Figs. 2c and d. In the paramagnetic state at 42 K, the nuclear reflections are identified at the (1, 0, 0) and (1, 1, 0) positions. Here, the former reflection is allowed only when the Co atoms form the regular triangular lattice as shown in Fig. 1c. In the Phase I at 10 K, additional magnetic reflections appear at the (1/2, 0, 0) and (1/2, 1/2, 0) positions, which suggests that the Phase I is characterized by the fundamental magnetic modulation vector ${\bf q} = (1/2, 0, 0)$.

To determine orientations of the local magnetic moments, we have further performed polarized neutron scattering experiments with the measurement configuration as shown in Fig. 2b. For polarized neutrons, Fourier-transformed spin components parallel and perpendicular to the neutron polarization direction ${\bf S}_n$ contributes to the non-spin-flip (NSF) and spin-flip (SF) scattering, respectively\cite{PolarizedNeutronTheory}. Here, we measured magnetic reflections on the $(H,K,0)$ scattering plane with an incident neutron beam polarized along the $c$ axis (${\bf S}_n \parallel c$), and collected intensities of the scattered neutrons with up spins by using a Heusler analyzer. The NSF and SF intensities were measured by changing the sign of the spin polarization of the incident neutrons using a spin flipper. In this configuration, NSF and SF scattering reflects the modulated spin components parallel and perpendicular to the $c$-axis, respectively. Figure 2f shows the scattering profile of magnetic reflection at the (1/2, 1/2, 0) position measured in the Phase I at 2.2 K near zero field. We found that both NSF and SF intensities are finite for this reflection, which clearly demonstrated that the magnetic structure in the Phase I contains both the out-of-plane ($\parallel c$) and in-plane ($\perp c$) spin components. It indicates that the previously proposed in-plane collinear or coplanar (i.e. non-collinear) magnetic structures\cite{ParkinNeutron, CoNb3S6_AHE, CoTa3S6_AHE, CrystalHall} must be reconsidered. We also measured the reflection at the (1/2, 0, 0) position as shown in Fig. 2e, and found that the NSF (SF) intensity is present (absent) for this reflection. These extinction rules are important to determine the symmetry of the magnetic structure as we discuss in the following.

To identify the possible magnetic structure from the viewpoint of symmetry, we have performed the representation analysis. In this approach, the exhaustive list of irreducible representations and associated basis vectors are obtained, whose linear combination can describe any magnetic structure symmetrically compatible with the given crystal structure and magnetic modulation vector ${\bf q}$\cite{CoM3S6_SuzukiCluster}. For each basis, the existence/absence of spontaneous Hall conductivity $\sigma_{xy} (=\rho_{yx}/(\rho_{xx}^2 + \rho_{yx}^2))$ can be judged from the Neumann's principle\cite{Conductivity_Symmetry}. In case of CoTa$_3$S$_6$ with ${\bf q} = (1/2, 0, 0)$, our analysis suggests that no single-${\bf q}$ magnetic order without net magnetization is allowed to host spontaneous Hall effect. On the other hand, we found that some triple-${\bf q}$ magnetic order can support the appearance of $\sigma_{xy} \neq 0$. Figure 3 indicates the list of triple-${\bf q}$ magnetic structure bases with $M=0$, which summarizes the magnetic point group (MPG), conductivity tensor $\sigma$, Fourier-transformed spin components ${\tilde {\bf m}}^{\perp}({\bf Q})$ for the wave vectors of (1/2, 0, 0) and (1/2, 1/2, 0), the corresponding NSF/SF neutron scattering channel, and the spin configuration for each basis. Here, intensities of magnetic scatterings are proportional to $|{\tilde {\bf m}}^{\perp}({\bf Q})|^2$, and the detailed definition of ${\tilde {\bf m}}^{\perp}({\bf Q})$ is provided in Methods section. Among them, the Bases 1 and 3 are characterized by the same magnetic point group ($62'2'$) as the uniform ferromagnetic state with $M \parallel c$, and the emergence of non-zero $\sigma_{xy}$ is allowed even for $M=0$. The Fourier components of the Basis 1 contains only the out-of-plane ($\parallel c$) spin component, and can contribute to the NSF scattering at both (1/2, 0, 0) and (1/2, 1/2, 0) positions, which agrees well with the experimental observation shown in Figs. 2e and f. On the other hand, the Basis 3 has only the in-plane ($\perp c$) spin component, and contributes to the SF scattering only at the (1/2, 0, 0) position, which obviously contradicts with the present results. This means that the Basis 3 is not included in the magnetic structure for Phase I. To account for the observed SF scattering, we need to add other triple-${\bf q}$ bases with in-plane spin components, namely the Bases 4-6. Among them, only the Basis 6 can contribute to the finite SF intensity at the (1/2, 1/2, 0) position, and it is also consistent with the absence of the SF intensity at the (1/2, 0, 0) position. Therefore, we conclude that the magnetic structure for Phase I consists of a linear combination of the Bases 1 and 6, as illustrated in Fig. 4a. This antiferromagnetic structure contains eight Co sites in a magnetic unit cell. When we focus on the green-shaded region of Fig. 4a, it can be viewed as the all-in-all-out type non-coplanar spin texture on two tetrahedral Co units. To confirm the validity of this magnetic sturcture, we have further performed unpolarized time-of-flight neutron Laue diffraction measurements, and collected intensities of 198 magnetic reflections (See Methods section and Supplementary Note III for the detail). Figure 4d indicates the comparison between the observed and calculated magnetic structure factors ($|F_{\rm cal}|$ and $|F_{\rm obs}|$) assuming the non-coplanar spin texture in Fig. 4a, which are in good agreement with each other with a reasonable reliability factor $R = 13 \%$. In Fig. 4e, $|F_{\rm cal}|$ and $|F_{\rm obs}|$ are plotted for the reflections on the $(1/2, 1/2, L)$ line. The calculation predicts that the magnetic scattering appears only for the even numbers of $L$, which is fully satisfied by the experimental data. The above results firmly establish the all-in-all-out noncoplanar antiferromagnetic structure (Fig. 4a) in Phase I.

To investigate the stability of such a non-coplanar spin texture, we have theoretically estimated the total energy of several antiferromagnetic and ferromagnetic spin orders for CoTa$_3$S$_6$ based on a first-principles DFT (density functional theory) calculation, as summarized in Supplementary Note II and Supplementary Fig. 2. Our calculation confirms that the all-in-all-out non-coplanar antiferromagnetic order (Fig. 4a) is characterized by the lower energy, i.e. more stable than the other collinear antiferromagnetic and ferromagnetic orders. In CoTa$_3$S$_6$, the ABAB type stacking of Co triangular lattice layers can be considered as the aggregates of local Co tetrahedra (Fig. 1c and Fig. 4a). While the inter-layer coupling is expected to be weak in intercalated van der Waals compounds, the intra-layer magnetic interaction between the neighboring Co ions is also mediated by the relatively weak super-super exchange interaction across two S sites. The tetrahedral lattice with antiferromagnetic interactions generally causes three-dimensional magnetic frustration\cite{Nagaosa_fcc, Blugel_TopologicalKerr_fcc, Ueda_Weyl, Ueda_AHE}, which may explain the appearance of non-coplanar spin order. On the other hand, recent theoretical studies predict that the itinerant-electron-mediated interactions can stabilize the all-in-all-out antiferromagnetic structure even for the single layer triangular lattice systems\cite{Batista, Motome_SSC_JPSJ, Motome_SSC}. Better intuitive understanding of the origin of the observed non-coplanar spin texture is the issue for the future study.

The all-in-all-out non-coplanar antiferromagnetic order (Fig. 4a) in CoTa$_3$S$_6$ is characterized by magnetic point group $32'$, which is the subgroup of that for ferromagnetic one ($62'2'$) with $M \parallel c$. It means that the associated conduction electrons should behave as if there is fictitious magnetic field along the $c$-axis, and the spontaneous Hall effect is allowed even without net magnetization. In such an environment, tiny additional spin canting is generally permitted by the Dzyaloshinskii-Moriya interaction\cite{weakFM_DM} and/or magneto-crystalline anisotropy\cite{weakFM_anisotropy}, which leads to the appearance of weak spontaneous magnetization $\Delta M$ along the $c$-axis. Note that the aforementioned spontaneous Hall effect is allowed even without this tiny spontaneous magnetization, since the all-in-all-out spin order itself breaks the time-reversal symmetry and hosts non-zero $\sigma_{xy}$. According to the Berry phase theory\cite{RMP_AHE, Nagaosa_fcc, Batista, Motome_SSC_JPSJ, Blugel_TopologicalKerr_fcc}, conduction electrons interacting with three non-coplanar spins ${\bf S}_i$, ${\bf S}_j$ and ${\bf S}_k$ on a triangular Co plaquette ($\alpha$) will feel a local fictitious magnetic flux ${\bf b}_\alpha \propto t_{\alpha} \chi_{\alpha} {\bf n}_\alpha$ (Fig. 1b and Fig. 4c). Here, $\chi_{\alpha} \equiv {\bf S}_i \cdot ({\bf S}_j \times {\bf S}_k)$ represents the scalar spin chirality scaling with the solid angle spanned by ${\bf S}_i$, ${\bf S}_j$ and ${\bf S}_k$, which is allowed to become non-zero only for the non-coplanar spin texture. $t_\alpha$ is the transfer integral along the loop $i \rightarrow j \rightarrow k \rightarrow i$, and ${\bf n}_\alpha$ is a unit vector normal to the triangular plaquette $\alpha$. When we focus on a single triangular lattice layer of Co atoms as shown in Fig. 4c, every triangular plaquette is characterized by the same sign of $\chi_{\alpha}$ and ${\bf b}_\alpha$. By taking a summation over a magnetic unit cell, it provides macroscopic fictitious magnetic field ${\bf b}_{\rm total} = \sum_{\alpha} {\bf b}_\alpha$ along the out-of-plane direction. Because of the short spin modulation period (i.e. large $\chi_\alpha$ value) and ferroic arrangement of ${\bf b}_\alpha$, the amplitude of ${\bf b}_{\rm total}$ can be very large. For CoTa$_3$S$_6$, the additional contribution can arise from triangular Co plaquettes containing the out-of-plane Co-Co bonds, while ${\bf b}_{\rm total}$ still survives along the $c$-direction due to the hexagonal symmetry of this compound (See Supplementary Note V for the detail)\cite{Nagaosa_fcc, Blugel_TopologicalKerr_fcc}. Therefore, the observed unconventionally large spontaneous Hall signal can be ascribed to the topological Hall effect (Fig. 1b), which originates from the fictitious magnetic field ${\bf b}_{\rm total}$ associated with the scalar spin chirality in non-coplanar antiferromagnetic orders.

Reflecting its time-reversal symmetry breaking, CoTa$_3$S$_6$ should host two degenerated antiferromagnetic domains (i.e. the domains A and B as shown in Figs. 4a and b) connected by time-reversal operation, which are characterized by the opposite sign of fictitious magnetic field, $\sigma_{xy}$, and $\Delta M$. By applying the external magnetic field $H \parallel c$, their degeneracy is lifted through $\Delta M$ and one of these domains can be selected. In this context, the hysteresis loop and the associated sign reversal of spontaneous Hall signal observed in Fig. 1g should represent the $H$-induced switching of time-reversal antiferromagnetic domains, where the domains A and B with the opposite sign of fictitious magnetic field are realized in the red and blue colored region, respectively. In case of cubic pyrochlore compounds with all-in-all-out spin order (that host ${\bf b}_{\rm total} = 0$ due to their cubic symmetry\cite{Nagaosa_fcc, Pyrochlore_Multipole}), the application of external magnetic field causes a transition into the 3-in-1-out spin state\cite{Ueda_Weyl}, and similar non-coplanar spin texture may be realized in Phase II above 4.5 T in CoTa$_3$S$_6$ (Figs. 1e-g) (See Supplementary Note V for the detailed discussion on the 3-in-1-out spin state).

In this study, we reported the experimental discovery of TRS-broken all-in-all-out non-coplanar antiferromagnetic order in triangular lattice compounds Co$M_3$S$_6$ ($M=$ Nb, Ta). Theoretically, the all-in-all-out spin texutre on triangular lattice has been considered as the most typical configuration to induce maximum amplitude of fictitious magnetic field associated with scalar spin chirality\cite{Nagaosa_fcc, Batista, Motome_SSC_JPSJ, Blugel_TopologicalKerr_fcc, Motome_SSC}. While its experimental realization has long been elusive, our findings suggest that intercalated van der Waals compounds can be an ideal model system to fulfill this situation. Combined with the observed unconventional electron transport properties, the present results evidence that scalar spin chirality mechanism can be another promising route to realize giant spontaneous Hall response in compensated antiferromagnets. 

In such systems, the fictitious magnetic field is predicted to cause not only topological Hall effect, but also various exotic phenomena such as topological magneto-optical effect\cite{Blugel_TopologicalKerr_fcc} and topological Nernst effect\cite{Gd2PdSi3_Nernst}, whose amplitude can be comparable or even larger than typical ferromagnets. The application of electric current is also expected to induce a driving force on TRS-broken spin texture through spin-transfer or spin-orbit torques\cite{Mn3Sn_Econtrol, Mn3Sn_Econtrol2}. They potentially enable nontrivial manner of efficient electrical and optical reading/writing of non-coplanar antiferromagnetic domains, and further exploration of similar material systems and their peculiar responses would be very attractive. The present compounds are also characterized by several unique additional features such as good exfoliatability into nanosheets\cite{CoNb3S6_AHE2} and noncentrosymmetric crystal structure, which may also contribute to the development of novel functionalities in TRS-broken antiferromagnets\cite{2DReviewMag2, NonreciprocalReview}.

\begin{figure}
\begin{center}
\includegraphics[width=13.5cm]{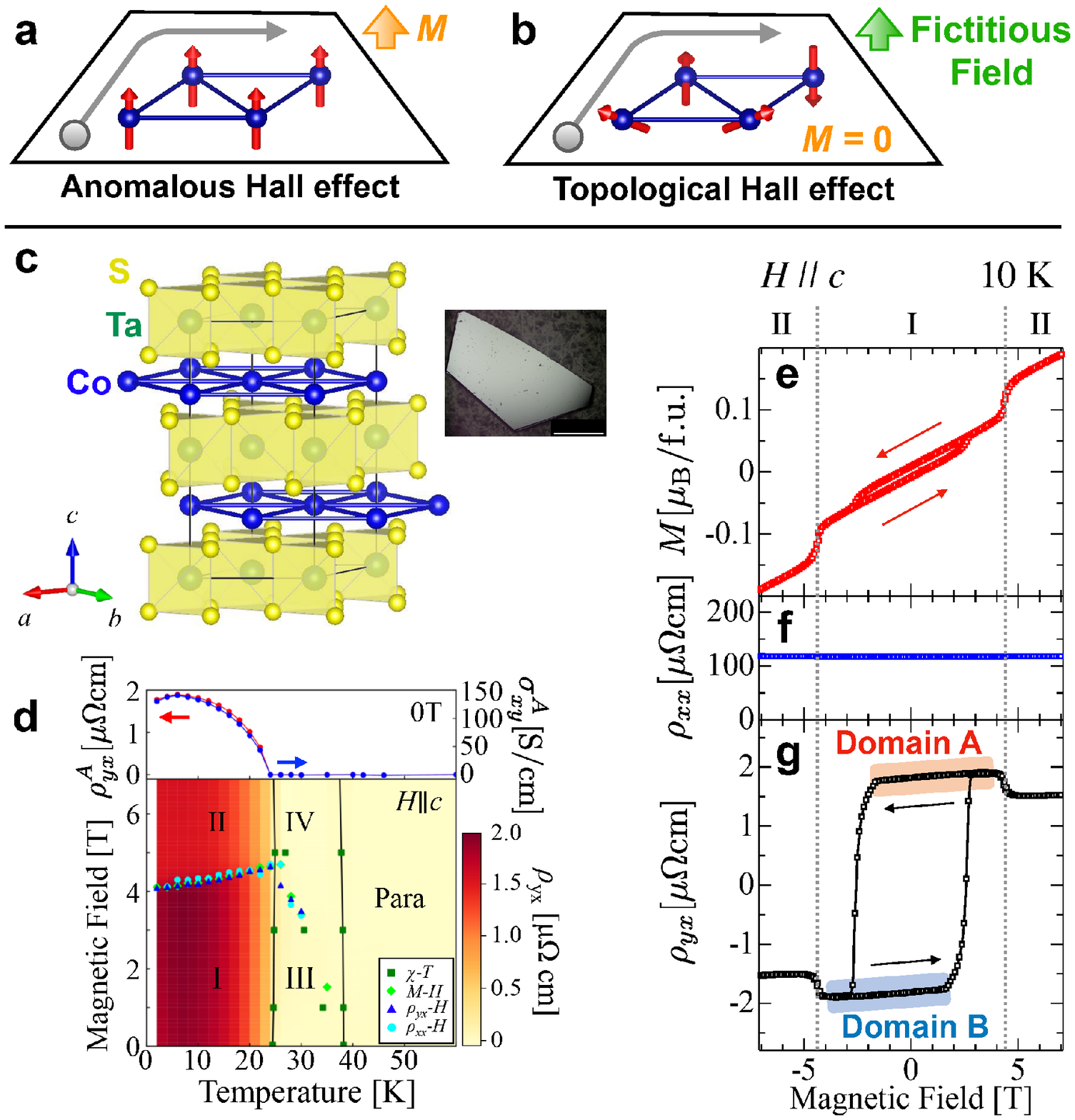}
\caption{{\bf Magnetic and electron transport properties of CoTa$_3$S$_6$.} {\bf a},{\bf b}, Schematic illustration of anomalous Hall effect and topological Hall effect. Gray circles indicate the charge carrier. {\bf c}, Crystal structure of CoTa$_{3}$S$_{6}$. The $[110]$, $[1\bar{1}0]$, and $[001]$ axes are defined as $x$, $y$ and $z$, respectively. Inset is a picture of an as-grown single crystal, where the white bar corresponds to 1 mm. {\bf d}, $H$(Magnetic field)-$T$(temperature) magnetic phase diagram for $H\parallel c$, determined based on the $H$ and $T$ dependence of magnetic susceptibility $\chi$, magnetization $M$, longitudinal resistivity $\rho_{\mathrm{xx}}$ and Hall resistivity $\rho_{\mathrm{yx}}$. The background color represents the amplitude of $\rho_{\mathrm{yx}}$. The upper panel represents the temperature dependence of spontaneous Hall resistivity and conductivity ($\rho_{xy}^A$ and $\sigma_{xy}^A$) at 0 T. {\bf e}-{\bf g}, Magnetic field dependence of $M$ ({\bf e}), $\rho_{\mathrm{xx}}$ ({\bf f}) and $\rho_{\mathrm{yx}}$ ({\bf g}) for $H \parallel c$ at 10K.}
\label{figure1}
\end{center}
\end{figure}

\begin{figure}
\begin{center}
\includegraphics[width=14.5cm]{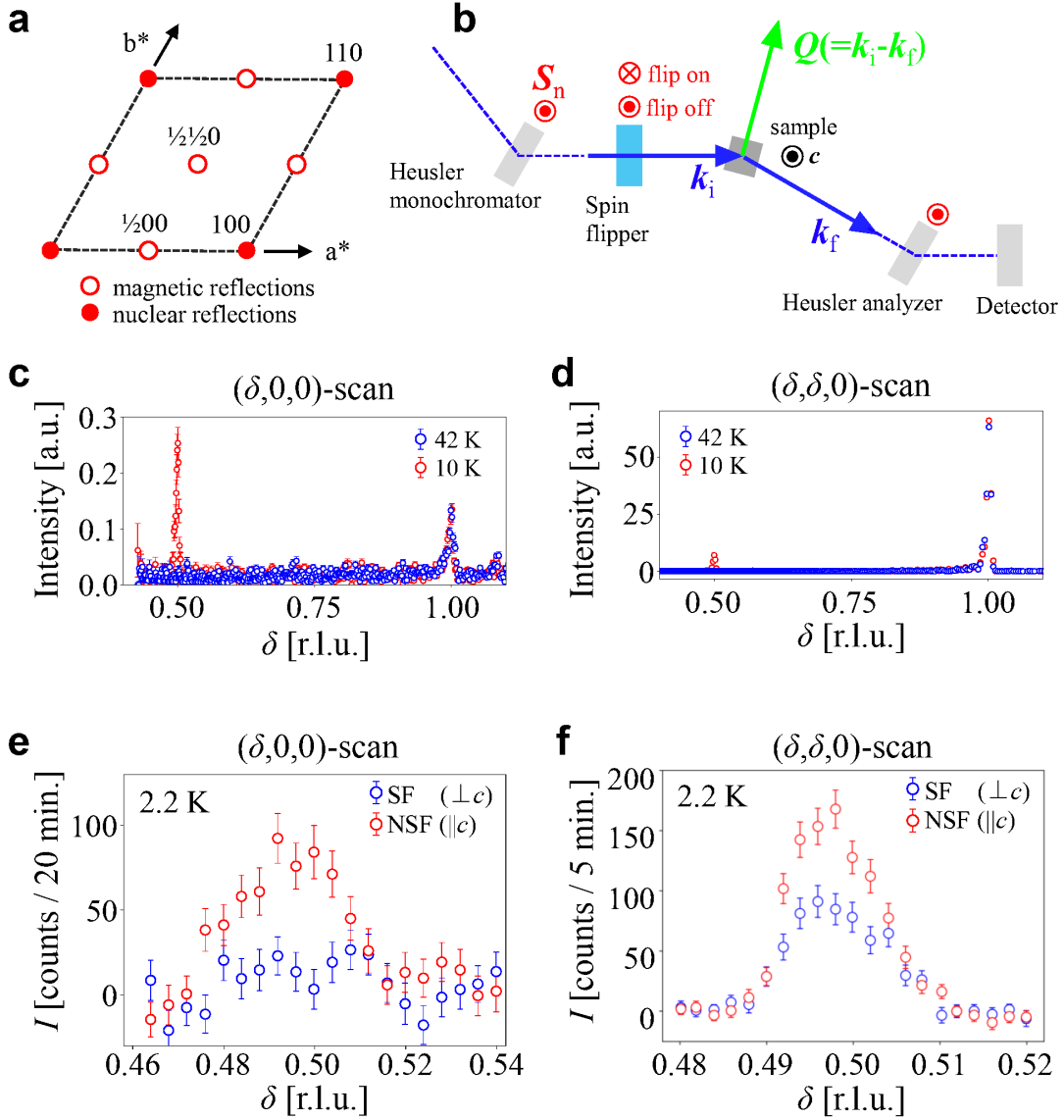}
\caption{{\bf Neutron scattering profiles in Phase I at $H=0$ for CoTa$_3$S$_6$.} {\bf a}, Schematic illustration of neutron diffraction patterns for the $(H, K, 0)$ plane in the reciprocal space for Phase I at $H=0$. The magnetic modulation vector is ${\bf q}=(1/2,0,0)$. {\bf b}, Measurement configuration for polarized neutron scattering experiments. {\bf c},{\bf d}, The ($\delta$, 0, 0) and ($\delta$, $\delta$, 0) line profiles of unpolarized neutron scattering intensity measured at 10 K (Phase I) and 42 K (paramagnetic state) with $H=0$. {\bf e},{\bf f}, The polarized neutron scattering profiles for the ($\delta$, 0, 0) and ($\delta$, $\delta$, 0) line scans measured in Phase I at 2.2 K near zero field. Intensities measured at 50 K in the paramagnetic phase are subtracted as background. Spin-flip (SF) and non-spin-flip (NSF) scattering represent the in-plane $(\perp c)$ and out-of-plane $(\parallel c)$ component of the modulated spin component, respectively. (See text for the detail.)}
\label{figure2}
\end{center}
\end{figure}

\begin{figure}
\begin{center}
\includegraphics[width=14.5cm]{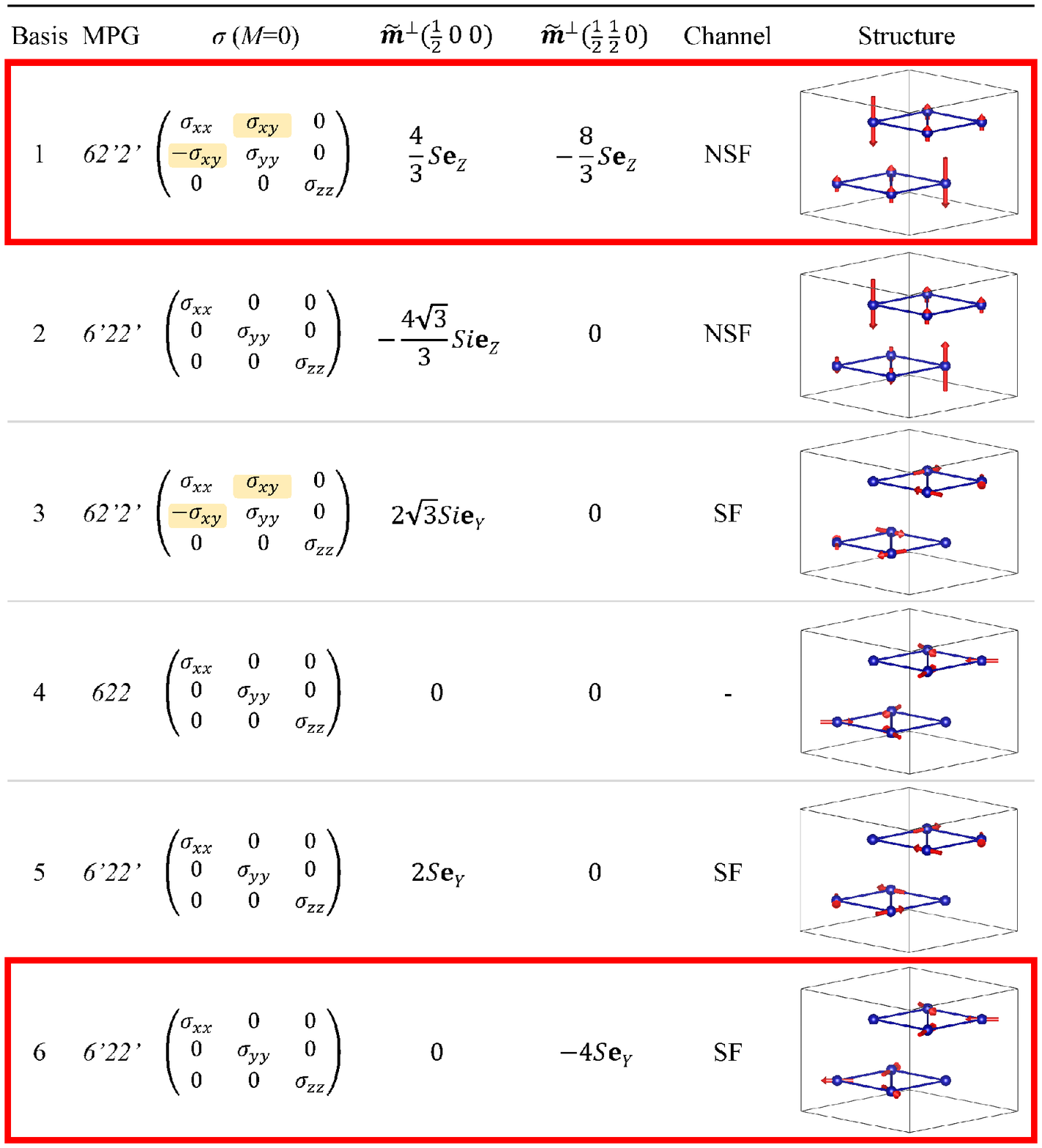}
\caption{{\bf List of triple-${\bf q}$ magnetic structure bases for CoTa$_3$S$_6$.} The triple-${\bf q}$ magnetic structure bases that satisfy ${\bf q}=(1/2, 0, 0)$ and $M=0$. For each basis, magnetic point group (MPG), conductivity tensor $\sigma$, Fourier-transformed spin component $\tilde{\bf m}^\perp({\bf Q})$ for the wave vectors ${\bf Q}=(1/2, 0, 0)$ and ${\bf Q}=(1/2, 1/2, 0)$, the corresponding NSF/SF neutron scattering channels, and the real-space spin arrangement are listed. ${\bf e}_Y$ and ${\bf e}_Z$ are unit vectors of a Cartesian coordinate system $X$, $Y$, $Z$, in which $X$ and $Z$ axes are defined to be parallel to the ${\bf Q}$-vector and the $c$ axis, respectively. $S$ represents the amplitude of magnetic moments denoted by the red arrows. Note that, for the Bases 1 and 2, the magnetic moments depicted by shorter arrows are defined to have the magnitudes of $S/3$. }
\label{figure3}
\end{center}
\end{figure}

\begin{figure}
\begin{center}
\includegraphics[width=16cm]{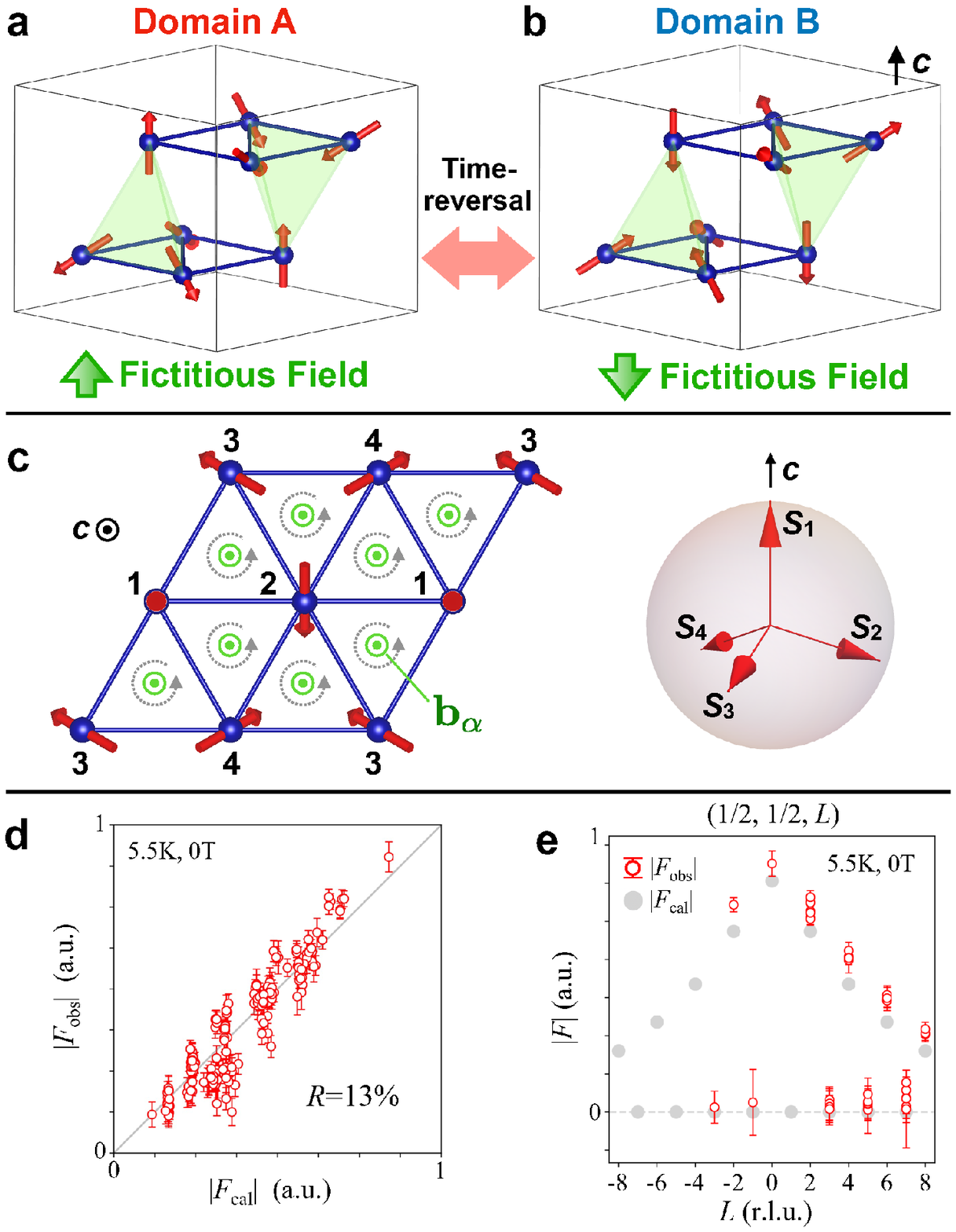}
\caption{{\bf All-in-all-out non-coplanar antiferromagnetic order in CoTa$_3$S$_6$ and its verification by neutron Laue diffraction experiments.} {\bf a},{\bf b}, Schematic illustration of the all-in-all-out non-coplanar antiferromagnetic order realized in Phase I, which is described by the linear combination of the Bases 1 and 6 in Fig. 3. There exist two possible time-reversal domains (i.e. the domains A and B), which are converted into each other by time-reversal operation. These two domains are characterized by the opposite sign of fictitious magnetic field along the $c$-axis. {\bf c}, Schematic illustration of a single triangular lattice layer of Co atoms in Phase I, which consists of four spin vectors ${\bf S}_n$ ($n=1, 2, 3, 4$) shown in the right panel. The dotted gray arrows indicate the direction of $i \rightarrow j \rightarrow k \rightarrow i$ loop for transfer integral $t_\alpha$. Green arrows pointing along the out-of-plane direction represent the local fictitious magnetic flux ${\bf b}_\alpha$. {\bf d}, Comparison between the calculated and observed magnetic structure factors ($|F_{\rm cal}|$ and $|F_{\rm obs}|$), assuming the all-in-all-out spin texture shown in {\bf a}. For this purpose, the intensities of 198 magnetic reflections are collected by performing unpolarized time-of-flight neutron Laue diffraction measurements (See Supplementary Note III for the detail). {\bf e}, $|F_{\rm cal}|$ and $|F_{\rm obs}|$ plotted for the ${\bf Q} =(1/2, 1/2, L)$ magnetic reflections.}
\label{figure3}
\end{center}
\end{figure}

\newpage

\section*{Methods}

\subsection*{Sample preparation and characterization.}
Polycrystalline sample of CoTa$_{3}$S$_{6}$ (CoNb$_{3}$S$_{6}$) was prepared from stoichiometric mixture of Co, Ta (Nb) and S sealed in a silica ampoule under vacuum. They were heated at 300℃ for 2 days and then at 850℃ for 1 day. Bulk single crystals were grown by chemical vapor transport using iodine as the transport agent. The polycrystalline powder (0.6 g) was loaded together with iodine (0.031 g) in a silica ampoule with inner diameters of 20 mm and 12.5 cm long. The ampoule was loaded in a horizontal tube furnace for 8 weeks (3 weeks) for CoTa$_3$S$_6$ (CoNb$_3$S$_6$), in which the temperature of the hot and cold zone were 950 $^\circ$C and 850 $^\circ$C, respectively. Crystal orientations were determined using the back-reflection X-ray Laue photography method, and the purity of the sample was confirmed by powder X-ray diffraction experiment. Note that the magnetic modulation vector reported in Ref. \cite{ParkinNeutron} for CoTa$_3$S$_6$  is different from the present work, and the possible origin of this discrepancy is discussed in Supplementary Note VI.

\subsection*{Magnetic and electrical transport properties measurements.}
Measurements of magnetic susceptibility and magnetization were performed using a superconducting quantum interference device magnetometer (MPMS, Quantum Design). Measurements of $\rho_{\mathrm{xx}}$ and $\rho_{\mathrm{yx}}$ were performed using the AC-transport option in a physical property measurement system (PPMS, Quantum Design). Here, we employed a standard 5-terminal method, and Au wires of 10 $\mu$m diameter were attached to the sample with silver paste as an electrode. An electric current of 10 mA and 117 Hz was used for the transport measurements. For consistency, the same single crystal piece was used for all measurements.\\

\subsection*{Neutron scattering experiments}

Neutron scattering measurements to find the magnetic modulation vector ${\bf q}$ were performed at a time-of-flight (TOF) small-angle and wide-angle neutron scattering instrument TAIKAN constructed at the BL15 of the Materials and Life Science Experimental Facility (MLF) of Japan Proton Accelerator Research Complex (J-PARC) in Japan\cite{TAIKAN}. %
A single crystal with dimensions of $5\times 5 \times 1$ mm$^3$ was used for the measurements. %
An incident neutron beam with a wavelength range from 0.7 to 7.7 \AA was exposed on the sample. %
The observed neutron intensities were processed by UTSUSEMI software\cite{UTSUSEMI}. %

Polarized neutron scattering measurements were performed at the polarized neutron triple-axis neutron spectrometer PONTA in JRR-3 of the Japan Atomic Energy Agency in Tokai Japan. %
A spin-polarized incident neutron beam was obtained by a Heusler(111) crystal monochromator. %
The neutron spin polarization was controlled by guiding fields, Helmholtz coil and a spin flipper. %
It should be noted that CoTa$_3$S$_6$ exhibits a weak ferromagnetism in the Phase I. %
To align the weak ferromagnetic domains, we applied a magnetic field of about 80 mT at the sample position using a pair of permanent magnets attached above and below the sample. %
We also employed a Heusler(111) crystal analyzer to analyze the spin state of the scattered neutrons. %
We measured the polarized neutron scattering profiles at 2.2 K in the Phase I and at 50 K in the paramagnetic phase. %
In Figs. 2e and f, we subtracted the intensities at 50 K from those at 2.2 K to remove nonmagnetic background signals. %
The total beam polarization was 0.86, and the effects of the imperfect polarization were corrected in Figs. 2e and f. %
To analyze the results of the polarized neutron scattering measurements, we calculated the Fourier-transformed magnetic moments ${\tilde {\bf m}}^{\perp}({\bf Q})$, which is given by %
\begin{eqnarray}
{\tilde {\bf m}}^{\perp}({\bf Q})=\sum_j{\bf S}^{\perp}_j\exp(i{\bf Q}\cdot{\bf d}_j).
\end{eqnarray}
where ${\bf S}^{\perp}_j$ and ${\bf d}_j$ are the spin moments projected onto a plane perpendicular to the scattering vector ${\bf Q}$ and the position vectors of $j$th Co atoms in a magnetic unit cell, respectively. Note that only the magnetic moment normal to the scattering vector ${\bf Q}$ generally contribute to the magnetic reflection of neutrons\cite{PolarizedNeutronTheory}. %
We listed ${\tilde {\bf m}}^{\perp}({\bf Q})$ for the possible models of triple-{\bf q} magnetic orders in Fig. 3. %
The nonzero Fourier components parallel and perpendicular to the $c$ axis result in NSF and SF scattering, respectively. 

TOF unpolarized neutron diffraction for the magnetic structure analysis was performed by using a single-crystal neutron diffractometer SENJU (BL18) in MLF of J-PARC\cite{SENJU}. %
Nuclear and magnetic reflections were collected at 5.5 K in zero field, and their integrated intensities were extracted by the program, STARGazer\cite{STARgazer}, taking into account the detector efficiencies, the wavelength dependence of the incident neutron flux and the absorption effect.
The crystal and magnetic structures were refined by JANA2006\cite{JANA} and original python program, respectively. %

\subsection*{Theoretical calculation of total energy for magnetically ordered states}

First-principles calculations based on the density functional theory were performed with the VASP package \cite{vasp1,vasp2,vasp3}.
The generalized gradient approximation proposed by Perdew, Burke and Ernzerhof was adopted for the exchange-correlation functional \cite{gga}.
We considered five types of collinear antiferromagnetic (AFM), ferromagnetic (FM), and all-in-all-out non-coplanar spin configurations shown in Supplementary Note II.
The lattice constant of collinear AFM, FM, and all-in-all-out non-coplanar structure is ($a=9.95$ \AA, $b=5.74$ \AA, $c=11.88$ \AA), ($a=b=5.74$ \AA, $c=11.88$ \AA), and ($a=b=11.48$ \AA, $c=11.88$ \AA), respectively. 
$10\times 12\times 8$, $12\times 12\times 5$, and $6\times 6\times 6$ $k$-point mesh with Monkhorst-Pack scheme \cite{MH} in the Brillouin zone sampling was used for AFM, FM, and all-in-all-out non-coplanar structure, respectively.
For the total-energy calculation, we confirmed the k-mesh dependence of the energy convergence in the FM structure.

\section*{Data availability} 

The data presented in the current study are available from the corresponding authors on reasonable request.

\section*{Author contributions} S.S., H.T. and R.T. planned the project. H.T., R.T., N.D.K., K.K., Y.T. and S.S. prepared the samples and performed the macroscopic measurements. H.T., T.N., S.S., R.K., K.O., H.S. and D.H. performed neutron and X-ray diffraction experiments. S.M., T.N., M.S., Y.Y., M.H. and R.A. performed the theoretical calculations. S.S. and H.T. wrote the manuscript with the support by T.N., S.M. and R.A. All authors discussed the results and commented on the manuscript.

\section*{Acknowledgements} The authors thank Y. Taguchi, T. Arima, S. Hayami, Y. Motome, N. Nagaosa, and S. Maekawa for enlightening discussions and experimental helps. This work was partly supported by Grants-In-Aid for Scientific Research (grant nos 18H03685, 19H01856, 19H05825, 20H00349, 20H05262, 20K05299, 20K21067, 21H01789, 21H04437, 21H04440, 21H04990, 21K13873, 21K13876, 21K18595) from JSPS, PRESTO (grant nos JPMJPR18L5, JPMJPR20B4, JPMJPR20L7) and CREST (grant no. JPMJCR1874) from JST, Katsu Research Encouragement Award of the University of Tokyo, Asahi Glass Foundation and Murata Science Foundation. The neutron scattering experiments at the Materials and Life Science Experimental Facility of the J-PARC and Japan Research Reactor 3 were performed under user programs (Proposal Nos. 2017L0701, 2020B0119, 21401, and 21511). The illustration of crystal structure was drawn by VESTA\cite{VESTA}.

\section*{Additional information} Supplementary information is available in the online version of the paper. Reprints and permissions information is available online at www.nature.com/reprints. Correspondence and requests for materials should be addressed to S.S. (email: seki@ap.t.u-tokyo.ac.jp)

\section*{Competing financial interests}  The authors declare that they have no competing financial interests.

\newpage

\end{document}


\title{Supplementary information:\\ Spontaneous topological Hall effect induced by non-coplanar antiferromagnetic order in intercalated van der Waals materials}

\author{H. Takagi$^{1}$, R. Takagi$^{1,2,3,4}$, S. Minami$^{5}$, T. Nomoto$^{6}$, K. Ohishi$^{7}$, M.-T. Suzuki$^{8,9}$, Y. Yanagi$^{8,10}$, M. Hirayama$^{1,4}$, N. D. Khanh$^{4}$, K. Karube$^{4}$, H. Saito$^{11}$, D. Hashizume$^{4}$, R. Kiyanagi$^{12}$, Y. Tokura$^{1, 4, 13}$, R. Arita$^{4,6}$, T. Nakajima$^{4,11}$, S. Seki$^{1,2,3,4}$}

\affiliation{$^1$ Department of Applied Physics, University of Tokyo, Tokyo 113-8656, Japan, \\ $^2$ Institute of Engineering Innovation, University of Tokyo, Tokyo 113-8656, Japan, \\ $^3$ PRESTO, Japan Science and Technology Agency (JST), Kawaguchi 332-0012, Japan, \\ $^4$ RIKEN Center for Emergent Matter Science (CEMS), Wako 351-0198, Japan, \\ $^5$ Department of Physics, University of Tokyo, Tokyo 113-8656, Japan, \\ $^6$ Research Center for Advanced Science and Technology, University of Tokyo, Tokyo 153-8904, Japan, \\ $^7$ Neutron Science and Technology Center, Comprehensive Research Organization for Science and Society (CROSS), Tokai 319-1106, Japan, \\ $^8$ Center for Computational Materials Science, Institute for Materials Research, Tohoku University, Sendai 980-8577, Japan, \\ $^9$ Center for Spintronics Research Network, Graduate School of Engineering Science, Osaka University, Toyonaka, Osaka 560-8531, Japan, \\ $^{10}$ Liberal Arts and Sciences, Toyama Prefectural University, Imizu, Toyama 939-0398, Japan, \\ $^{11}$ The Institute for Solid State Physics, University of Tokyo, Kashiwa, Chiba, Japan, \\ $^{12}$ J-PARC Center, Japan Atomic Energy Agency, Tokai, Naka-gun 319-1195, Japan, \\ $^{13}$ Tokyo College, University of Tokyo, Tokyo 113-8656, Japan.}

\maketitle

\section{Magnetic field dependence of magnetization and Hall resistivity measured at various temperatures for CoTa$_3$S$_6$}

In this section, we discuss the fundamental magnetic and electron transport properties for CoTa$_3$S$_6$. The upper panel of Supplementary Fig. 1a shows the temperature dependence of magnetic susceptibility $\chi$ measured at 1 T for $H \parallel c$. It is characterized by two kinks at $T_{{\rm N1}}$ = 38 {\rm K} and $T_{{\rm N2}}$ = 24 {\rm K}, which suggests the appearance of two distinctive antiferromagnetic phases. In the lower panel of Supplementary Fig. 1a, temperature dependence of magnetization $M$ measured under zero magnetic field after the field cooling at 1 {\rm T} is also indicated, where the appearance of tiny spontaneous magnetization ($\Delta M \sim 0.01 \mu_B/$Co) along the $c$-axis has been identified below $T_{{\rm N2}}$. Such a spontaneous $M$ is confirmed to be absent along the direction normal to the $c$-axis. 

\begin{figure}
\begin{center}
\includegraphics*[width=14.5cm]{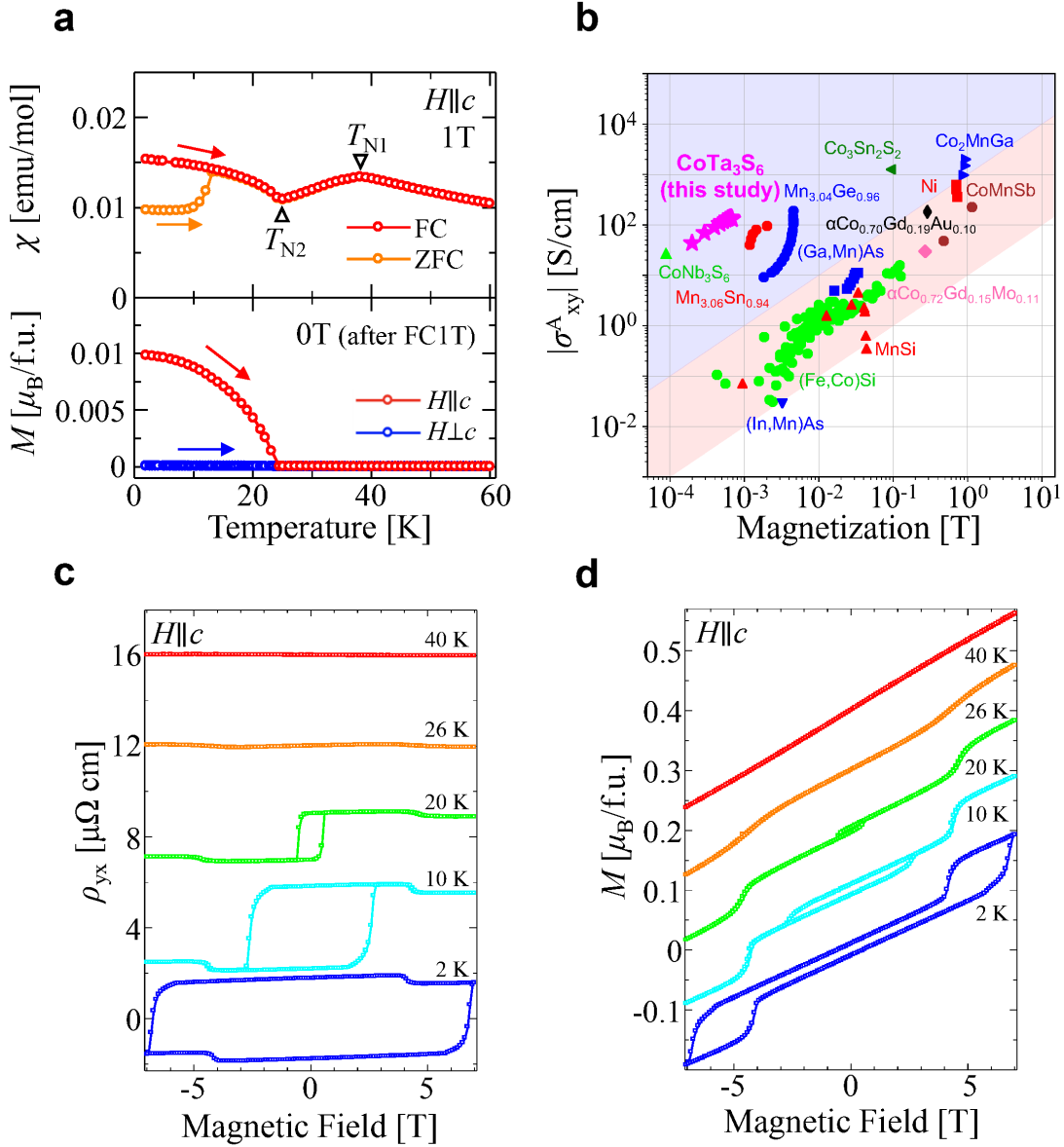}
\caption{{\bf Magnetic and electron transport properties of CoTa$_3$S$_6$.} {\bf a}, Upper panel: Temperature dependence of magnetic susceptibility $\chi$ for $H \parallel c$, measured at 1 T after the field cooling(FC) and zero field cooling(ZFC). Lower panel: Temperature dependence of magnetization $M$ at 0 T, measured after the field-cooling at 1 T. {\bf b}, Logarithmic plot of the magnetization dependence of the amplitude of spontaneous Hall conductivity $|\sigma_{xy}^A|$ for CoTa$_{3}$S$_{6}$ and CoNb$_3$S$_6$\cite{CoNb3S6_AHE}, and other previously reported magnetic materials cited from Ref. \cite{Mn3Sn_Review}. The red shades indicate the region for conventional ferromagnets, where $|\sigma^{\rm A}_{xy}|$ is almost linearly proportional to magnetization. CoTa$_{3}$S$_{6}$ and CoNb$_3$S$_6$ are located in the blue region, where exceptionally large $|\sigma^{\rm A}_{xy}|$ is observed. {\bf c},{\bf d}, Magnetic filed dependence of Hall resistivity $\rho_{\mathrm{yx}}$ ({\bf c}) and magnetization $M$ ({\bf d}) measured with $H\parallel c$ at various temperatures. Each data is shifted along the vertical direction for clarify.}
\label{figure1}
\end{center}
\end{figure}

In Supplementary Fig. 1c and d, magnetic field dependence of $\rho_{yx}$ and $M$ measured for $H \parallel c$ at various temperatures are plotted. The detailed discussion on the data at 10 K is provided in the main text. Below $T_{\mathrm{N2}} \sim 24$ {\rm K}, $\rho_{\mathrm{yx}}$ and $M$ profiles are always characterized by the spontaneous Hall/magnetization signal with clear hysteresis loop, as well as a step-like anomaly at 4.5 {\rm T}. At lower temperatures, the critical magnetic field value required for the sign-reversal of $\rho_{yx}$ and $M$ is gradually enhanced. Above $T_{\mathrm{N2}}$, the spontaneous Hall and magnetization components and associated hysteresis loop disappear, and only a small anomaly corresponding to another metamagnetic transition is observed up to $T_{\mathrm{N1}} \sim 38$ K. These behaviors are consistent with the previous reports in Ref. \cite{CoTa3S6_AHE}, except for the slight difference of critical field value possibly dependent on the defect density in the sample.

On the basis of these $M$, $\rho_{\mathrm{xx}}$ and $\rho_{\mathrm{yx}}$ profiles, $H$-$T$ magnetic phase diagram for $H\parallel c$ is summarized in Fig. 1d in the main text, where four distinctive magnetic phases (I, II, III and IV) are identified. The anomalies in $M$, $\rho_{\mathrm{xx}}$ and $\rho_{\mathrm{yx}}$ profiles always coincide with each other, which indicates the strong correlation between magnetism and electron transport properties. Here, the background color represents the amplitude of $\rho_{\mathrm{yx}}$ in the field decreasing process. It suggests that the clear enhancement of $\rho_{\mathrm{yx}}$ is observed in the Phases I and II. The upper panel of Fig. 1d in the main text indicates the temperature dependence of spontaneous Hall resistivity $\rho^{\mathrm{A}}_{\mathrm{yx}}$ (i.e. the $\rho_{\mathrm{yx}}$ value at $H=0$ in Supplementary Fig. 1c) and the corresponding spontaneous Hall conductivity $\sigma^{\mathrm{A}}_{xy}$ (i.e. the value of $\sigma_{\mathrm{xy}}=\rho_{\rm{yx}} /(\rho_{\rm xx}^2 + \rho_{\rm yx}^2)$ at $H=0$). These data confirm that the spontaneous Hall effect appears only below $T_{\mathrm{N2}} \sim 24$ K. In Supplementary Fig. 1b, the amplitude of spontaneous Hall conductivity $|\sigma^{\mathrm{A}}_{\mathrm{xy}}|$ is plotted against spontaneous magnetization $\Delta M$ for CoTa$_{3}$S$_{6}$ and CoNb$_3$S$_6$\cite{CoNb3S6_AHE} as well as the other previously reported magnetic compounds\cite{Mn3Sn_Review}. Here, conventional ferromagnets are located in the red shaded region, where $|\sigma^{\mathrm{A}}_{\mathrm{xy}}|$ almost linearly scales with $\Delta M$. On the other hand, a series of recently reported magnetic Weyl semimetals, such as antiferromagnetic Mn$_{3}X$ ($X$=Sn,Ge)\cite{Mn3Sn_Nature, Mn3Ge} and ferromagnetic Co$_2$MnGa\cite{Co2MnGa} and Co$_3$Sn$_2$S$_2$\cite{Co3SnS2}, are located in the blue region characterized by exceptionally large amplitude of $|\sigma^{\mathrm{A}}_{\mathrm{xy}}|$. Interestingly, the present CoTa$_{3}$S$_{6}$ and CoNb$_3$S$_6$ are also located in the latter blue region, which turn out to host the largest $|\sigma^{\mathrm{A}}_{\mathrm{xy}}| / \Delta M$ ratio ever reported. This strongly suggests the nontrivial origin of spontaneous Hall effect in CoTa$_{3}$S$_{6}$ and CoNb$_3$S$_6$.

\section{Theoretical calculation of total energy for all-in-all-out and other magnetic structures}

\begin{figure}[b]
\begin{center}
\includegraphics*[width=\linewidth]{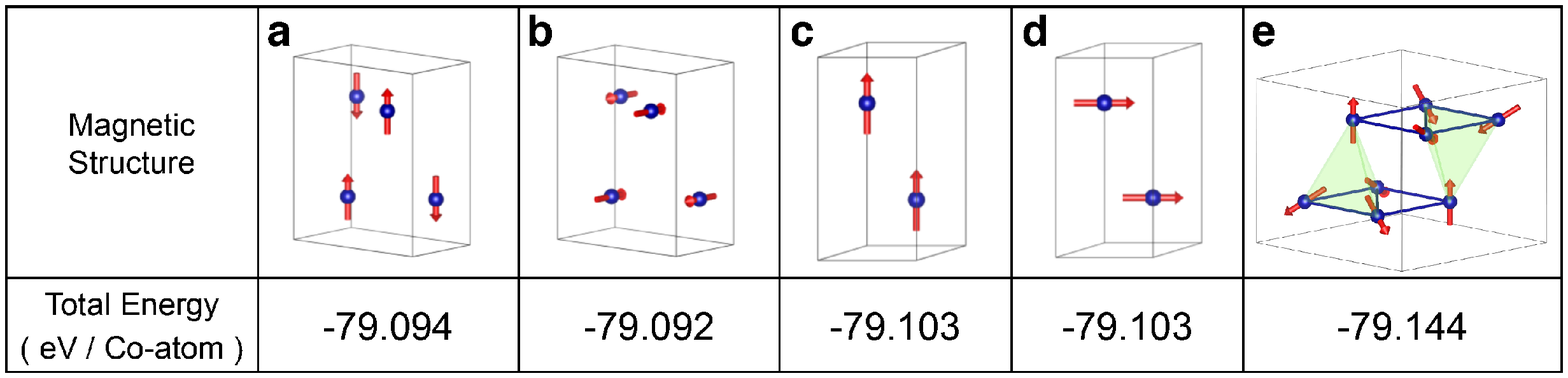}
\caption{{\bf Theoretical calculation of total energy for various magnetic structures for CoTa$_3$S$_6$.}
Total energy of collinear antiferromagnetic states ({\bf a},{\bf b}), ferromagnetic states ({\bf c},{\bf d}) and all-in-all-out non-coplanar antiferromagnetic state ({\bf e}) for CoTa$_3$S$_6$, estimated based on the first-principles DFT calculation. (See Methods section in the main text for the detail.)}
\end{center}
\label{figS2}
\end{figure}

To investigate the stability of the all-in-all-out non-coplanar antiferromagnetic order (Fig. 4a in the main text), the total energy calculations were carried out for CoTa$_3$S$_6$ with various spin arrangements based on density functional theory (DFT) (See Methods section in the main text for the detail). Here, we considered collinear antiferromagnetic states (Supplementary Figs. 2a and b), ferromagnetic states (Supplementary Figs. 2c and d), and all-in-all-out non-coplanar antiferromagnetic state (Supplementary Fig. 2e). Their magnetic unit cell contains four, two and eight Co ions, respectively, and the approximately the same density of $k$-mesh is chosen for each case. The total energy for these spin textures are summarized in Supplementary Fig. 2. It indicates that the all-in-all-out non-coplanar antiferromagnetic order is characterized by lower energy, i.e. more stable than the other collinear antiferromagnetic and ferromagnetic orders. These calculation results are consistent with the experimental observation in the main text.

\section{Magnetic structure analysis for CoTa$_3$S$_6$ based on neutron Laue diffraction}

As discussed in the main text, the results of the present polarized neutron scattering experiments suggest the appearance of all-in-all-out non-coplanar antiferromagnetic structure (Fig. 4a in the main text) in Phase I for CoTa$_3$S$_6$. To verify this structure, we have further performed time-of-flight (TOF) unpolarized neutron Laue diffraction measurements.

\begin{figure}[b]
\begin{center}
\includegraphics*[width=7cm]{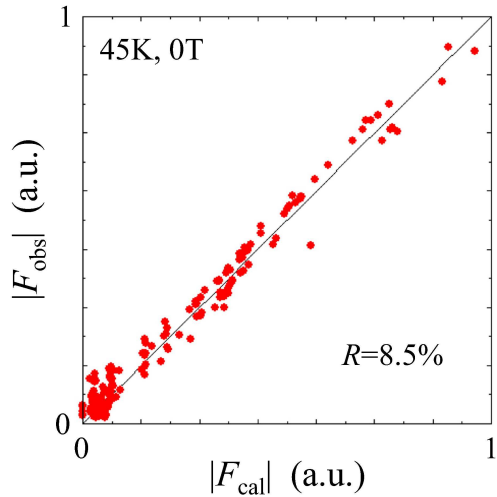}
\caption{{\bf Crystal structure analysis of CoTa$_3$S$_6$ by neutron Laue diffraction experiments.} Comparison between the calculated and observed nuclear structure factors ($|F_{\rm cal}|$ and $|F_{\rm obs}|$), for the crystal structure summarized in Supplementary Table 1. The measurements were performed in the paramagnetic state at 45 K and 0 T for CoTa$_3$S$_6$.}
\end{center}
\end{figure}

The experiment was carried out at SENJU diffractometer in MLF of J-PARC (see Methods section in the main text).
The intensities of magnetic and nuclear reflections were collected without external magnetic field. The integrated intensities were converted into structure factors taking into account Lorentz correction, sensitivity of the detectors, and wavelength-dependences of the incident neutron flux and neutron absorption of the sample. 

First, we performed crystal structure analysis using the nuclear structure factors. Supplementary Fig. 3 shows a comparison between observed and calculated nuclear structure factors. The refined structural parameters are summarized in Supplementary Table 1. Here, we assumed that Ta and S sites are fully occupied, and refined only the occupancy ($g$) of the Co site. As a result, we obtained $g=0.92(2)$ for the Co site. We also checked molar ratio of Co to Ta by means of the inductively coupled plasma atomic emission spectroscopy (ICP-AES) analysis, revealing that the ratio is Co : Ta = 0.90 : 3. This is consistent with the present crystal structure analysis by neutron diffraction.  

\begin{table*}[b]
\begin{center}
\begin{tabular}{cccccc}
\hline
\hline
 & site & $g$ & $x$ & $y$  & $z$\\
\hline
Co & $2c$ & 0.92(2) & $\frac{1}{3}$ & $\frac{2}{3}$ & $\frac{1}{4}$ \\
Ta1 & $4f$ & 1 & $\frac{1}{3}$ & $\frac{2}{3}$ & $0.5009(1)$  \\
Ta2 & $2a$ & 1 & 0 & 0 & 0 \\
S & $12i$ & 1 & 0.3345(5) & 0.0006(5) & 0.1316(1) \\
\hline
\hline
\end{tabular}
\caption{Structural parameters of CoTa$_3$S$_6$, obtained by the crystal structure analysis based on the experimental data in Supplementary Fig. 3. The space group is $P6_3 22$ and the lattice parameters are $a=5.74$ \AA ~and $c=11.93$ \AA.}
\label{Table_S1}
\end{center}
\end{table*}

We then performed magnetic structure analysis in Phase I at 5.5 K. The scale factors between the observed and calculated structure factors were fixed at the values obtained by the crystal structure analysis performed at the same temperature. 
We employed the magnetic form factor of Co$^{2+}$ for the calculation of the magnetic structure factors, although the valence state of Co in this system has not been identified yet. %
As discussed in the main text, the all-in-all-out non-coplanar spin texture (Fig. 4a in the main text) is described by the linear combination of the Bases 1 and 6 in Fig. 3 in the main text. Here, the orientation of local magnetic moment ${\bf m}_i$ at the site $i(=1, 2, 3, 4)$ of each Co tetrahedron is described by
\begin{eqnarray}
{\bf m}_1 &=& a_1{\bf e}_z\\
{\bf m}_2 &=& a_6{\bf e}_y - \frac{a_1}{3}{\bf e}_z\\
{\bf m}_3 &=& a_6 \left ( \frac{\sqrt{3}}{2}{\bf e}_x -\frac{1}{2}{\bf e}_y \right )  - \frac{a_1}{3}{\bf e}_z\\
{\bf m}_4 &=& a_6 \left (- \frac{\sqrt{3}}{2}{\bf e}_x -\frac{1}{2}{\bf e}_y \right )  - \frac{a_1}{3}{\bf e}_z.
\end{eqnarray}
${\bf e}_x$, ${\bf e}_y$, and ${\bf e}_z$ are the unit vectors along the $[110]$, $[1\bar{1}0]$, and $[001]$ axes, respectively. %
$a_1$ and $a_6$ represent the amplitude of the Bases 1 and 6, respectively. %
We refined these amplitudes by comparing the magnetic structure factors calculated from the above model with the observed ones, and the best fit to the data is obtained for $a_1 =1.6 \mu_{\rm B}/$Co and $a_6 =0.78 \mu_{\rm B}/$Co. %
Figure 4d in the main text indicates the comparison between the observed and calculated magnetic structure factors ($|F_{\rm obs}|$ and $|F_{\rm cal}|$), which are in good agreement with each other with a reasonable reliability factor $R = 13 \%$. %
We also confirmed that the ratio between the SF and NSF intensities in the polarized neutron scattering experiment can be reproduced by the amplitudes of the two bases above, as discussed in Supplementary Note IV. %

In Fig. 4e in the main text, $|F_{\rm obs}|$ and $|F_{\rm cal}|$ are plotted for the reflections on the $(0.5, 0.5, L)$ line. The calculation predicts that the magnetic scattering appears only for the even numbers of $L$. This reflection condition is fully satisfied by the experimental data. 

On the basis of the above results and the polarized neutron scattering results in the main text, we concluded that the all-in-all-out non-coplanar spin texture is realized in Phase I.

\section{Polarized neutron scattering experiments for CoNb$_3$S$_6$}

\begin{figure}[b]
\begin{center}
\includegraphics*[width=16.5cm]{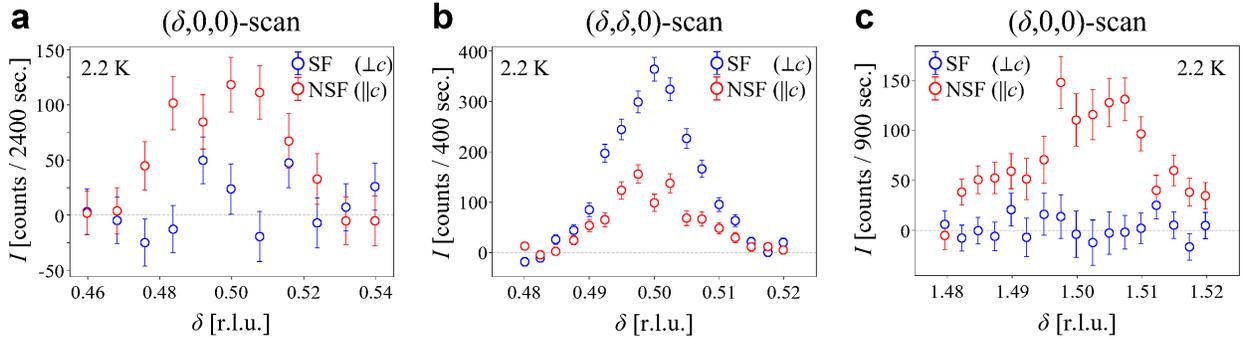}
\caption{{\bf Polarized neutron scattering profiles for CoNb$_3$S$_6$.} {\bf a}-{\bf c}, The polarized neutron scattering profiles for the ($\delta$, 0, 0) and ($\delta$, $\delta$, 0) line scans measured at 2.2 K near zero field. {\bf a}, {\bf b}, and {\bf c} represent the magnetic reflections at the (1/2, 0, 0), (1/2, 1/2, 0), and (3/2, 0, 0) positions, respectively. The measurement was performed with the configuration shown in Fig. 2b in the main text. Spin-flip (SF) and non-spin-flip (NSF) scattering represent the in-plane $(\perp c)$ and out-of-plane $(\parallel c)$ component of the modulated spin component, respectively. (See text for the detail.)}
\end{center}
\end{figure}

\begin{figure}[b]
\begin{center}
\includegraphics*[width=15cm]{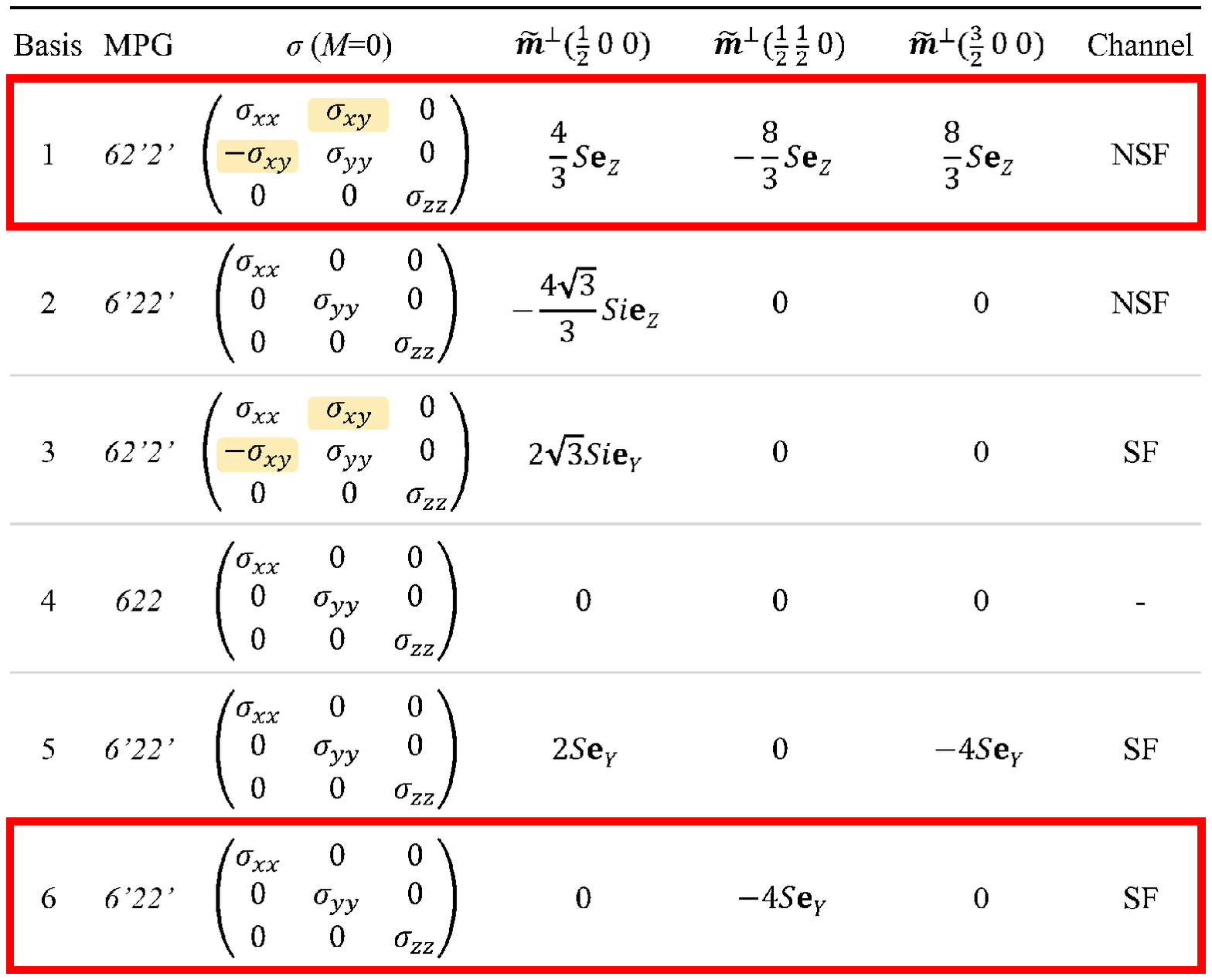}
\caption{{\bf List of triple-${\bf q}$ magnetic structure bases for CoNb$_3$S$_6$.} The triple-${\bf q}$ magnetic structure bases that satisfy ${\bf q}=(0.5, 0, 0)$ and $M=0$ for CoNb$_3$S$_6$. For each basis, magnetic point group (MPG), conductivity tensor $\sigma$, Fourier-transformed spin component $\tilde{\bf m}^\perp({\bf Q})$ for the wave vectors ${\bf Q}=(1/2, 0, 0)$, ${\bf Q}=(1/2, 1/2, 0)$ and ${\bf Q}=(3/2, 0, 0)$, and the corresponding NSF/SF neutron scattering channels are listed. The label for each basis is common with Fig. 3 in the main text, in which the corresponding real-space spin texture is also presented. ${\bf e}_Y$ and ${\bf e}_Z$ are unit vectors of a Cartesian coordinate system $X$, $Y$, $Z$, in which $X$ and $Z$ axes are defined to be parallel to the ${\bf Q}$-vector and the $c$ axis, respectively.}
\end{center}
\end{figure}

In this section, we discuss the results of polarized neutron scattering experiments for CoNb$_3$S$_6$. Below $T_{\rm N} \sim 28$ K, this compound was reported to host large spontaneous Hall effect despite its tiny spontaneous magnetization along the $c$-axis ($\Delta M \sim 0.0013 \mu_{\rm B}$/Co)\cite{CoNb3S6_AHE}, and the reported $M$ and $\rho_{yx}$ profiles are well reproduced in our CoNb$_3$S$_6$ crystal. Previous unpolarized neutron scattering experiment reported the magnetic modulation vector ${\bf q} = (1/2, 0, 0)$, and a collinear antiferromagnetic order has been proposed for this compound\cite{ParkinNeutron, CrystalHall}. 

The measurement configuration for polarized neutron scattering experiment is the same as the one employed for CoTa$_3$S$_6$ (Fig. 2b in the main text), where the SF and NSF scattering reflect the modulated spin components perpendicular and parallel to the $c$-axis, respectively. The measurement was performed near zero field at 2.2 K. In Supplementary Fig. 4b, the results for magnetic reflection at the (1/2, 1/2, 0) position are indicated. Since both SF and NSF scattering intensities are clearly identified, the magnetic structure should contain both the in-plane ($\perp c$)  and out-of-plane ($\parallel c$) spin components. To identify the possible magnetic structure from the viewpoint of symmetry, we have performed the representation analysis following the procedures as detailed in the main text. In Supplementary Figs. 4a-c, the polarized neutron scattering results are presented for the (1/2, 0, 0), (1/2, 1/2, 0), and (3/2, 0, 0) positions. Here, the finite NSF scattering intensity is observed for all three reflections, while SF scattering intensity is observed only for the (1/2, 1/2, 0) position. Supplementary Fig. 5 indicates the list of possible triple-${\bf q}$ magnetic structure bases with $M=0$, which covers the same information as Fig. 3 in the main text, as well as the Fourier-transformed spin component $\tilde{{\bf m}}^\perp ({\bf Q})$ for the wave vector ${\bf Q}=(3/2, 0, 0)$. By considering the existence of spontaneous Hall effect (i.e. $\sigma_{xy} \neq 0$) in CoNb$_3$S$_6$\cite{CoNb3S6_AHE} and the aforementioned selection rules of polarized neutron scattering in Supplementary Figs. 4a-c, we concluded that the magnetic structure of CoNb$_3$S$_6$ is described by the linear combination of the Bases 1 and 6 as in case of CoTa$_3$S$_6$, which represents the non-coplanar all-in-all-out antiferromagnetic order (Supplementary Fig. 6b). 

\begin{figure}[t]
\begin{center}
\includegraphics*[width=14cm]{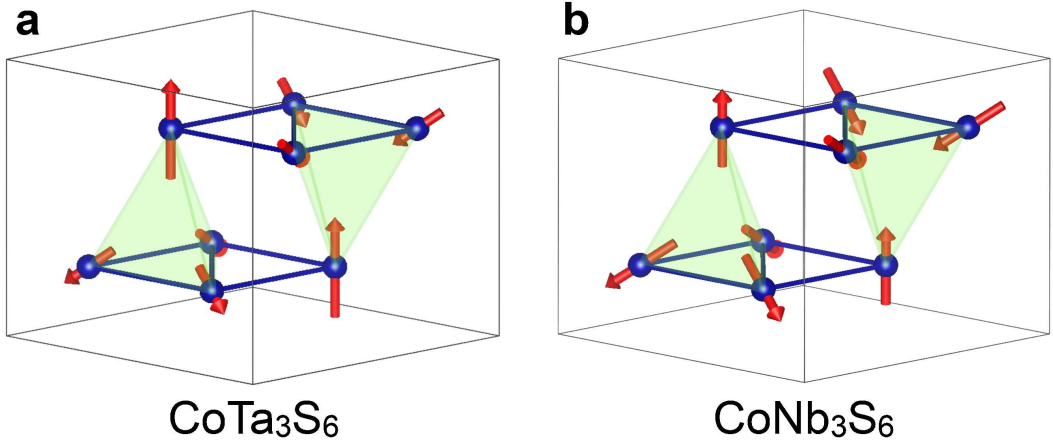}
\caption{{\bf All-in-all-out non-coplanar magnetic structure in CoTa$_3$S$_6$ and CoNb$_3$S$_6$.} {\bf a},{\bf b}, Schematic illustration of magnetic structure for CoTa$_3$S$_6$ and CoNb$_3$S$_6$ in the magnetic ground state at $H=0$, determined based on the polarized neutron scattering experiments.}
\end{center}
\end{figure}

The observed selection rule of polarized neutron scattering is common between CoNb$_3$S$_6$ and CoTa$_3$S$_6$, while these two compounds host different intensity ratio between the SF and NSF scattering at the (1/2, 1/2, 0) position (Fig. 2f in the main text and Supplementary Fig. 4b). From these data, the $a_6/a_1$ ratio (i.e. the linear combination ratio of the Bases 1 and 6 as introduced in Supplementary Note III) is estimated to be 0.49 and 1.05 for CoTa$_3$S$_6$ and CoNb$_3$S$_6$, respectively. The former value is consistent with the one independently obtained from the neutron Laue diffraction experiments ($a_6/a_1 = 0.49$) in Supplementary Note III. The corresponding spin textures for CoTa$_3$S$_6$ and CoNb$_3$S$_6$ are illustrated in Supplementary Fig. 6a and b, respectively.

\section{Scalar spin chirality and topological Hall effect in CoTa$_3$S$_6$ and CoNb$_3$S$_6$}

\begin{figure}[b]
\begin{center}
\includegraphics*[width=15cm]{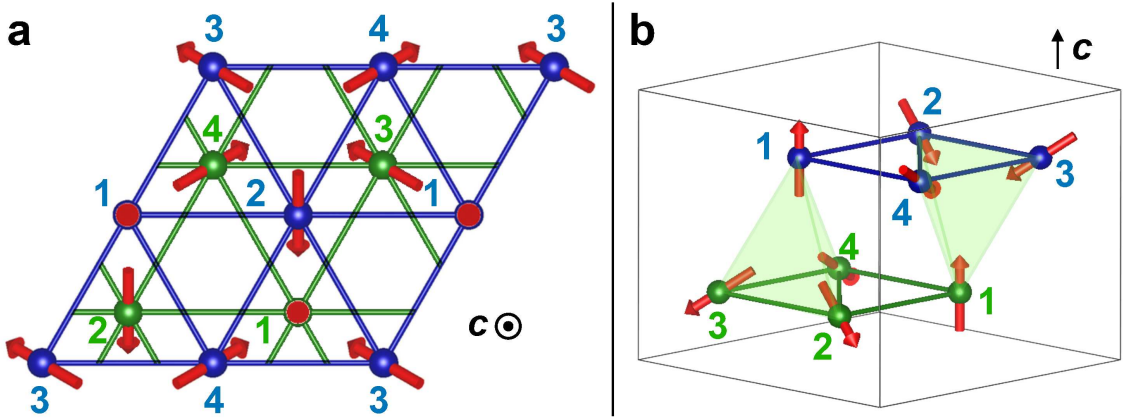}
\caption{{\bf Magnetic unit cell of all-in-all-out non-coplanar antiferromagnetic order in Co$M_3$S$_6$.} {\bf a},{\bf b}, Top view (${\bf a}$) and side view (${\bf b}$) of all-in-all-out non-coplanar spin order in Co$M_3$S$_6$. Here, blue and green circles indicate the Co atoms on the first and second triangular-lattice layer, respectively. Eight Co atoms are contained in a magnetic unit cell, which is characterized by four independent spin vectors labeled as 1-4.}
\end{center}
\end{figure}

\begin{figure}[b]
\begin{center}
\includegraphics*[width=15cm]{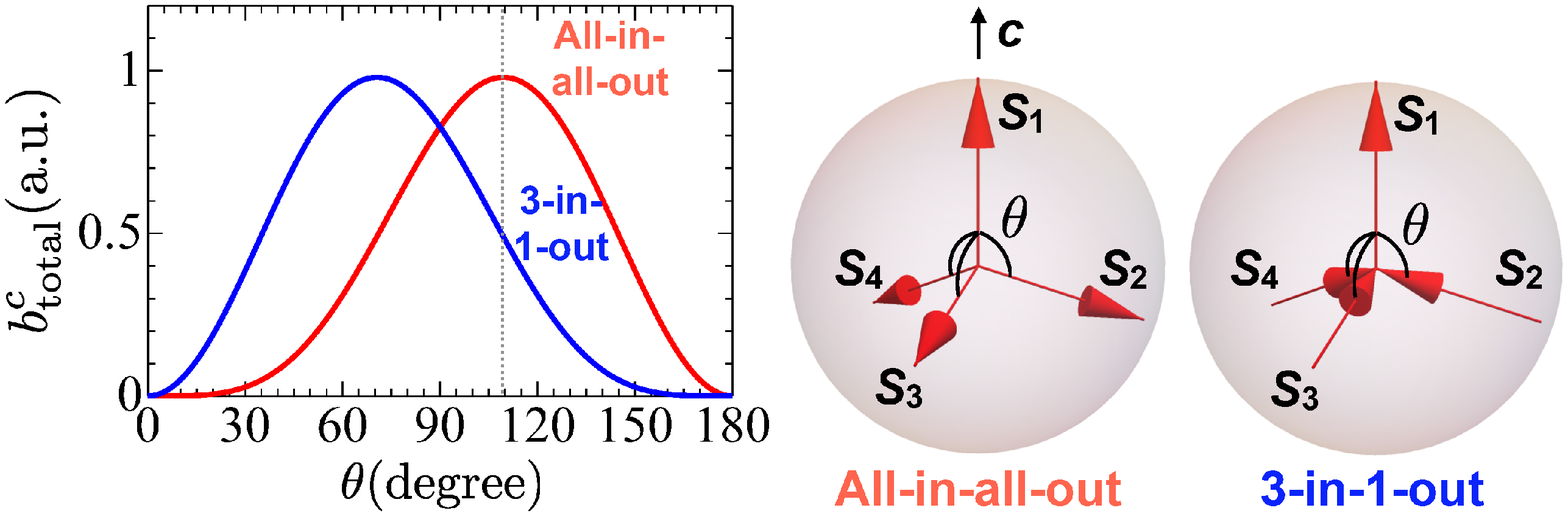}
\caption{{\bf $\theta$-dependence of fictitious magnetic field for the all-in-all-out and 3-in-1-out spin states in Co$M_3$S$_6$.} $b^c_{\rm total}$ indicates the $c$-component of macroscopic fictitious magnetic field ${\bf b}_{\rm total}$. All-in-all-out spin order is assumed to consist of four independent spin vectors ${\bf S}_1$, ${\bf S}_2$, ${\bf S}_3$, and ${\bf S}_4$ described by Eq. (S5)-(S8). $\theta$ is defined as the angle between ${\bf S}_1$ and the other ${\bf S}_n$ $(n=2,3,4)$ vectors. At $\theta= \arccos (-1/3) \sim 109.5^\circ$, these spin vectors form a regular tetrahedron if they are parallel transported to have a common origin. 3-in-1-out spin order is obtained by reversing the sign of ${\bf S}_2$, ${\bf S}_3$ and ${\bf S}_4$.}
\end{center}
\end{figure}

According to the Berry phase theory\cite{Nagaosa_fcc, RMP_AHE, Blugel_TopologicalKerr_fcc, Taguchi}, conduction electrons interacting with three non-coplanar spins ${\bf S}_i$, ${\bf S}_j$ and ${\bf S}_k$ on a triangular Co plaquette ($\alpha$) will feel a local fictitious magnetic flux ${\bf b}_\alpha \propto t_{\alpha} \chi_{\alpha} {\bf n}_\alpha$ (Fig. 1b and Fig. 4c in the main text). Here, $\chi_{\alpha} \equiv {\bf S}_i \cdot ({\bf S}_j \times {\bf S}_k)$ represents the scalar spin chirality scaling with the solid angle spanned by ${\bf S}_i$, ${\bf S}_j$ and ${\bf S}_k$, which is allowed to become non-zero only for the non-coplanar spin texture. $t_\alpha$ is the transfer integral along the loop $i \rightarrow j \rightarrow k \rightarrow i$, and ${\bf n}_\alpha$ is a unit vector normal to the triangular plaquette $\alpha$.

In the following, we assume the all-in-all-out non-coplanar spin texture consisting of four normalized spin vectors
\begin{eqnarray}
{\bf S}_1 &=& {\bf e}_z\\
{\bf S}_2 &=& {\bf e}_y \sin \theta + {\bf e}_z \cos \theta \\
{\bf S}_3 &=& {\bf e}_x \frac{\sqrt{3}}{2} \sin \theta - {\bf e}_y \frac{1}{2} \sin \theta + {\bf e}_z \cos \theta\\
{\bf S}_4 &=& -{\bf e}_x \frac{\sqrt{3}}{2} \sin \theta - {\bf e}_y \frac{1}{2} \sin \theta + {\bf e}_z \cos \theta,
\end{eqnarray}
where $\theta$ is defined as the angle between ${\bf S}_1$ and the other ${\bf S}_n$ $(n=2,3,4)$ vectors. At $\theta= \arccos (-1/3) \sim 109.5^\circ$, these vectors form a regular tetrahedron if they are parallel transported to have a common origin (Middle panel of Supplementary Fig. 8). In case of a single triangular lattice layer of Co atoms as shown in Fig. 4c in the main text, every triangular plaquette consisting of in-plane Co-Co bonds ($\alpha \in \alpha_{\rm intra}$) is characterized by the same sign of $\chi_{\alpha}$ and ${\bf b}_\alpha$. By taking a summation over a magnetic unit cell, it provides macroscopic fictitious magnetic field ${\bf b}_{\rm intra} \equiv \sum_{\alpha \in \alpha_{\rm intra}} {\bf b}_\alpha$ along the out-of-plane direction. For Co$M_3$S$_6$ with the ABAB-type stacking of triangular lattice Co layers (Supplementary Fig. 7), we have to consider additional triangular plaquettes containing the out-of-plane Co-Co bonds ($\alpha \in \alpha_{\rm inter}$) and associated ${\bf b}_{\rm inter} \equiv \sum_{\alpha \in \alpha_{\rm inter}} {\bf b}_\alpha$. In such a situation, the relationship ${\bf b}_{\rm inter} = - \gamma {\bf b}_{\rm intra}$ with $\gamma = \kappa \sqrt{3/(4\delta^2 -1)}$ generally holds, where $\delta$ is defined as the ratio of the out-of-plane Co-Co bond length to the in-plane one, and $\kappa$ is the ratio of $t_\alpha$ value for $\alpha \in \alpha_{\rm inter}$ to the one for $\alpha \in \alpha_{\rm intra}$. Since Co$M_3$S$_6$ should host $\delta \neq 1$, $\kappa \neq 1$ and $\gamma \neq 1$, finite amplitude of ${\bf b}_{\rm total} = {\bf b}_{\rm inter} + {\bf b}_{\rm intra}$ survives along the $c$-direction\cite{Nagaosa_fcc, Blugel_TopologicalKerr_fcc}. In this context, the unconventionally large spontaneous Hall signal observed in Co$M_3$S$_6$ can be ascribed to the topological Hall effect (Fig. 1b in the main text), which originates from the fictitious magnetic field ${\bf b}_{\rm total}$ associated with the scalar spin chirality in non-coplanar antiferromagnetic orders.

In case of cubic pyrochlore lattice compounds with all-in-all-out spin order (that hosts ${\bf b}_{\rm total} = 0$ due to its cubic symmetry), the application of external magnetic field is reported to induce a transition into the 3-in-1-out spin state\cite{Ueda_Weyl}. For the present Co$M_3$S$_6$, the 3-in-1-out spin order is obtained by reversing the sign of ${\bf S}_2$, ${\bf S}_3$ and ${\bf S}_4$ in Eq. (S5)-(S8). Supplementary Fig. 8 indicates the $\theta$-dependence of fictitious magnetic field $b^c_{\rm total}$ (i.e. the $c$-component of ${\bf b}_{\rm total}$) for the all-in-all-out and 3-in-1-out spin states in Co$M_3$S$_6$. At $\theta \sim 109.5^\circ$, 3-in-1-out spin texture is characterized by the same sign but smaller amplitude of $b^c_{\rm total}$ compared with the all-in-all-out spin texture. This behavior is consistent with the abrupt decrease of $|\rho_{yx}|$ upon the transition from Phase I to Phase II in Fig. 1g in the main text, when we assume the all-in-all-out spin order in Phase I and the 3-in-1-out spin order in Phase II. For the quantitative analysis of the observed $\rho_{yx}$-$H$ profile, the experimental identification of $\theta$ value in the latter state and the further consideration of $M$-linear anomalous Hall term would be necessary, which is the issue for the future study.

\section{Magnetic modulation vector in CoTa$_3$S$_6$}

As discussed in the main text and Supplementary Note I, the CoTa$_3$S$_6$ crystal used in the present study hosts two-step magnetic transitions as a function of temperature, where the lower-$T$ magnetic phase (i.e. Phase I) is characterized by weak spontaneous magnetization along the $c$-axis, large spontaneous Hall signal, and the magnetic modulation vector ${\bf q} = (1/2, 0, 0)$. Its magnetization and Hall resistivity profiles are in accord with the ones reported in Ref. \cite{CoTa3S6_AHE}. In these two works, the single crystalline samples were prepared by the chemical vapor transport method with similar temperature conditions, where the temperature at the hot and cold zones were 950 $^\circ$C and 850 $^\circ$C (940 $^\circ$C and 860 $^\circ$C) in the present work (Ref. \cite{CoTa3S6_AHE}), respectively. For the present CoTa$_3$S$_6$ crystal, the additional crystal structure analysis by neutron Laue diffractometer (Supplementary Note III), as well as the clear identification of nuclear reflection at the (1, 0, 0) position (Fig. 2c in the main text), confirms the $P6_3 22$ crystal structure with the $\sqrt {3} \times \sqrt {3}$ ordering of Co sites as shown in Fig. 1c in the main text.

On the other hand, Ref. \cite{ParkinNeutron} reported the magnetic modulation vector ${\bf q} = (1/3, 1/3, 0)$ for CoTa$_3$S$_6$, while its sample shows only a single magnetic phase below $T_{\rm N}$ with no spontaneous magnetization along the $c$-axis\cite{ParkinMagnetization}. In that paper, the detailed conditions for single crystal growth were not provided, and we naively expect that the difference of  growth temperature may be the origin of this discrepancy.

Note that the isostructural CoNb$_3$S$_6$ is also characterized by the magnetic modulation vector ${\bf q} = (1/2, 0, 0)$\cite{ParkinNeutron}. Since both CoNb$_3$S$_6$ and CoTa$_3$S$_6$ are confirmed to host similar spontaneous Hall effect, magnetic modulation vector ${\bf q} = (1/2, 0, 0)$, and all-in-all-out non-coplanar antiferromagnetic order in the present study, we concluded that the observed features must be intrinsic. In Ref. \cite{CoTa3S6_AHE}, the spontaneous Hall effect in CoTa$_3$S$_6$ was interpreted by assuming the coplanar spin texture based on the previously reported ${\bf q} = (1/3, 1/3, 0)$. Our present polarized neutron scattering results indicate that the spontaneous Hall effect rather originates from non-coplanar spin texture with ${\bf q} = (1/2, 0, 0)$, and its detailed mechanism is discussed in the main text.